\documentclass[11pt]{iopart}
\usepackage{iopams}
\usepackage{graphicx}
\usepackage{graphics}

\begin{document}
\title{On the behaviour of short Kratky-Porod chain}
\author{Semjon Stepanow}
\address{Martin-Luther-Universit\"{a}t Halle, Fachbereich Physik, D-06099 Halle,
Germany}
\date{\today }
\ead{stepanow@physik.uni-halle.de}

\begin{abstract}
Using the exact computation of a large number of moments of the
distribution function of the end-to-end distance $G(r,N)$ of the
worm-like chain, we have established the analytical form of the
coefficients in Taylor expansions of the moments for short chain lengths
$N$. The knowledge of these coefficients enabled us to resummate the
moment expansion of $G(r,N)$ by taking into account consecutively the
deviations of the moments from their stiff rod limit. Within this
procedure we have derived the short-chain expansion for $G(r,N)$, the
scattering function, and the extension-force relation, which take into
account the deviations of the moments from their stiff rod limit to the
seventh order in $N$.
\end{abstract}

\footnotetext{Dedicated to the 60th birthday of Professor Lothar
Sch\"{a}fer} \maketitle

\section{Introduction}

Polymers with contour length $L$ much larger than the persistence length $%
l_{p}$, which is the correlation length for the tangent-tangent correlation
function along the polymer and is quantitative measure of the polymer
stiffness, are flexible and are described by using the tools of quantum
mechanics and quantum field theory \cite{edwards65}-\cite{schaefer-buch}. If
the chain length decreases, the chain stiffness becomes an important factor.
Many polymer molecules have internal stiffness and cannot be modelled by the
model of flexible polymers developed by Edwards \cite{edwards65}.

The standard coarse-graining model of a worm-like polymer was proposed by
Kratky and Porod \cite{kratky/porod49}. The essential ingredients of this
model is the penalty for bending energy, and the local inextensibility.
The latter make the treatment of the model much more difficult. There is
a substantial amount of studies of the Kratky-Porod model in the last
half century \cite{hermans/ullman52}-\cite{yamakawa} (and the references
therein). In recent years there is an increasing interest in theoretical
description of semiflexible polymers
\cite{wilhelm/frey96}-\cite{stepanow04} (and the references therein). A
reason for this interest is due to potential applications of semiflexible
polymers in biology and in researches on semi-crystalline polymers.\

In this article we present results of the study of the behaviour of the
worm-like chain for short lengths. The consideration is based on our recent
work \cite{SS02},\cite{stepanow04}, where the Fourier-Laplace transform
$G(k,p)$ of the end-to-end distribution function $G(r,N)$ ($p$ is the Laplace
conjugate to $N$) was represented as the matrix
element of the infinite order matrix $\tilde{P}(k,p)=(I+ikDM)^{-1}D$ with
matrices $D$ and $M$ related to the spectrum of the quantum rigid rotator. A
truncation of $\tilde{P}(k,p)$ by matrix of order $n$ gives the end-to-end
distribution function $G(k,p)$ as a rational function being an infinite
series in powers of $k^{2}$, i.e. it contains all moments of the end-to-end
distribution function, and describes the first $2n-2$ moments exactly. In
context of eigenstates of the quantum rigid rotator, the truncation at order
$n$ takes into account the eigenstates with quantum number of the angular
momentum up to the value $l=n-1$. The moment expansion of $G(k,p)$ can be
represented as double series in powers of $k^{2}$ and $1/p$. Alternatively,
one can expand $G(k,N)$ in double series in powers of $k^{2}$ and $N$.
Analyzing the series of known moments (we have analytically calculated the
first $50$ moments), we have established that the coefficients in subseries
in powers of $k^{2}$ have a simple analytical structure, which enables one
to perform a partial resummation of the moment expansion of $G(k,p)$ or $%
G(k,N)$ (see below). This resummation procedure results in an expansion of $%
G(r,N)$ around the stiff rod limit, where the end-to-end distribution
function is given by the expression, $G_{\mathrm{r}}(r,N)=1/(4\pi
N^{2})\delta (N-r)$, so that because $G_{\mathrm{r}}(r,N)$ is a
distribution, so derived short-chain expansion is not a Taylor expansion
but rather an expansion in the space of distributions. The knowledge of
$G(k,N)$ enables one to compute directly the scattering function $S(k,N)$
and the extension-force relation\ $R(f)$. In this paper we present the
results of the calculation of terms of the short-chain expansions of
$G(r,N)$, $S(k,N)$, and\ $R(f)$ by taking into account the deviations of
the moments from their stiff rod behaviour to the seven order in chain
length $N$. The procedure can be extended to higher orders. The
short-chain expansions besides the interest in their own can be used for
comparisons with approximative treatments, and also in studies of the
behaviour of short semiflexible polymers.

The present paper is organized as follows. Section \ref{formalism}
introduces to the description of the worm-like chain using the formalism
of quantum rigid rotator. Section \ref{short-ch} explains the idea of the
derivation of the short-chain expansion, and presents the short-chain
expansions for the distribution function of the end-to-end distance,
scattering function, and extension-force relation.

\section{The formalism}

\label{formalism}

The Fourier transform of the distribution function of the end-to-end polymer
distance of the continuous Kratky-Porod model \cite{kratky/porod49} $G(%
\mathbf{k},L)=\int d^{3}R\exp (-i\mathbf{k(\mathbf{R}-\mathbf{R}_{0})})G(%
\mathbf{R}-\mathbf{R}_{0},L)$ is expressed by the path integral as follows
\begin{equation}
\fl G(\mathbf{k},L)=\int D\mathbf{t}(s)\prod\limits_{s}\delta (\mathbf{t}%
(s)^{2}-1)\exp (-i\mathbf{k}\int_{0}^{L}ds\mathbf{t}(s)-\frac{l_{p}}{2}%
\int_{0}^{L}ds(\frac{d \mathbf{t}(s)}{d s})^{2}),  \label{w1}
\end{equation}%
where $l_{p}$ is the persistence length, and $\mathbf{t}(s)=d \mathbf{%
r}(s)/d s$ is the tangent vector at the point $s$ along the contour of
the polymer. The product over $s$ in Eq.(\ref{w1}) takes into account
that the polymer chain is locally inextensible. In the following the arc
length of the polymer $L$ will be measured in units of $l_p$ and will be
denoted by $N$. We now will consider the Green's function $P(\theta
,\varphi ,N;\theta _{0},\varphi _{0},0)$
associated with Eq.(\ref{w1}). The differential equation for $P$ is%
\begin{equation}
\frac{\partial }{\partial N}P(\theta ,\varphi ,N;\theta _{0},\varphi _{0},0)-%
\frac{1}{2}\nabla _{\theta ,\varphi }^{2}P+U(\Omega )P=\delta (N)\delta
(\Omega -\Omega _{0}),  \label{w4}
\end{equation}%
where $U(\mathbf{kt}_{\Omega })=i\mathbf{k\mathbf{t}_{\Omega }}$ is the
potential energy of the rigid rotator in an external field $i\mathbf{k}$,
where $\mathbf{k}$ is measured in units of $l_{p}^{-1}$. The end-to-end
distribution function is obtained from $P(\Omega ,N;\Omega _{0},0)$ as
follows
\begin{equation}
G(k,N)=\frac{1}{4\pi }\int d\Omega \int d\Omega _{0}P(\Omega ,N;\Omega
_{0},0).  \label{w6}
\end{equation}%
The differential equation (\ref{w4}) can be rewritten as an integral
equation as follows%
\begin{equation}
\fl P(\Omega ,N;\Omega _{0},0)=P_{0}(\Omega ,N;\Omega _{0},0)-\int_{0}^{N}ds\int
d\Omega ^{\prime }P_{0}(\Omega ,N;\Omega ^{\prime },s)U(\mathbf{kt}_{\Omega
^{\prime }})P(\Omega ^{\prime },s;\Omega _{0},0),  \label{w5}
\end{equation}%
where the bare Green's function $P_{0}(\theta ,\varphi ,N;\theta
_{0},\varphi _{0},0)$ reads
\begin{equation}
P_{0}(\theta ,\varphi ,N;\theta _{0},\varphi _{0},0)=\sum_{l,m}\exp (-\frac{%
l(l+1)N}{2})Y_{lm}(\theta ,\varphi )Y_{lm}^{\ast }(\theta _{0},\varphi _{0}),
\label{w3}
\end{equation}%
with $Y_{lm}(\theta ,\varphi )$ being the spherical harmonics, and $l$ and $%
m $ are the quantum numbers of the angular momentum. Due to the
convolution character of the expression (\ref{w5}) with respect to the
integration over the contour length ($P_{0}(\Omega ,N;\Omega ^{\prime
},s)$ depends on the difference $N-s$), the Laplace transform of
$P(\Omega ,N;\Omega _{0},0)$ in Eq.(\ref{w5}) with respect to $N$ permits
to get rid of integrations over the contour length. Thus, in the
following we will consider the Laplace transform of $G(k,N)$ with respect
to $N$.

It was shown in \cite{SS02},\cite{stepanow04} that the solution of
Eq.(\ref{w5}) results in the following expression of the Fourier-Laplace
transform of the end-to-end distribution function as the matrix element
of an infinite order square matrix
\begin{equation}
G(k,p)=\left\langle 0\mid \tilde{P}^{s}(k,p)\mid 0\right\rangle  \label{w8a}
\end{equation}%
with
\begin{equation}
\tilde{P}^{s}(k,p)=(I+ikDM^{s})^{-1}D,  \label{w9a}
\end{equation}%
where the square matrices $M^{s}$ and $D$ are defined by%
\begin{equation}
M_{l,l^{\prime }}^{s}=w_{l}\delta _{l,l^{\prime }+1}+w_{l+1}\delta
_{l+1,l^{\prime }},  \label{w11}
\end{equation}%
with $w_{l}=\sqrt{l^{2}/(4l^{2}-1)}$, and
\begin{equation}
D_{l,l^{\prime }}=\frac{1}{\frac{1}{2}l(l+1)+p}\delta _{l,l^{\prime }},
\label{w10}
\end{equation}%
respectively. The superscript $s$ specifies that the quantities
$\tilde{P}^{s}$ and $M^{s}$ are square matrices. The quantity $\langle
0\mid \tilde{P}^{s}(k,p)\mid 0\rangle$ denotes the $(1,1)$ matrix element
of the infinite order square matrix $\tilde{P}^{s}$. Since the $(1,1)$
matrix element corresponds to the expectation value of the quantum rigid
rotator in the ground state with the quantum number $l=0$, we prefer to
use the above notation. Summations over the magnetic quantum number in
the intermediate states in the expression of $G(k,p)$ obtained by solving
Eq.(\ref{w5}) can be eliminated \cite{stepanow04}, so that summations in
the intermediate states occur only over the eigenvalues of the angular
momentum $l=0,1,...$. This is the reason why the calculation of $G(k,p)$
reduces to the computation of the matrix element of an infinite order
square matrix.

The quantity $\tilde{P}^{s}(k,p)$ plays the key role in the theory
similar to the bare propagator in common quantum field theories. The
end-to-end
distribution function $G(k,p)$ is simply the matrix element $\left\langle 0|%
\tilde{P}^{s}(k,p)|0\right\rangle $, the scattering function of the
polymer is the inverse Laplace transform of $G(k,p)/p^{2}$ multiplied by
$2/N$, the partition function of the stretched polymer is
$Z(f,N)=G(k=-if/k_{B}T,N)$ etc \cite{stepanow04}.

\section{The behaviour of short worm-like polymer}

\label{short-ch}

\subsection{The idea of the short-chain expansion}

To derive the short-chain expansion for the Kratky-Porod chain we insert
the
expansion of $D$ given by Eq.(\ref{w10}) according to%
\begin{equation*}
\frac{1}{\frac{1}{2}l(l+1)+p}=\frac{1}{p}\sum_{m=0}^{\infty }\frac{%
(-1)^{m}\left( l(l+1)\right) ^{m}}{2^{m}}\frac{1}{p^{m}}
\end{equation*}%
into the moment expansion of \ $G(k,p)$%
\begin{equation}
G(k,p)=\frac{1}{p}\sum_{m=0}^{\infty }(-1)^{m}\left\langle 0\mid
(DM^{s})^{2m}\mid 0\right\rangle (k^{2})^{m},  \label{mom-exp}
\end{equation}%
and obtain as a result $G(k,p)$ as double series in powers of ($k/p)^{2}$
and $1/p$ as follows%
\begin{equation}
G(k,p)=\sum_{s=1}^{\infty }\ \frac{1}{p^{s}}\sum_{n=0}^{\infty
}C_{n}^{s}\left( \frac{k^{2}}{p^{2}}\right) ^{n}  \label{w12}
\end{equation}%
The analysis of $50$ exactly computed moments of $G(r,p)$, which are
proportional to the coefficients $\left\langle 0\mid (DM^{s})^{2m}\mid
0\right\rangle$ in Eq.(\ref{mom-exp}),  shows that the coefficients
$C_{n}^{s}$ are polynomials in powers of $n$ of the order $2s-3$, i.e.
they have the form $C_{n}^{s}=a_{1}n+a_{2}n^{2}+...+a_{2s-3}n^{2s-3}$
($s\geq 2$). We determine the coefficients $a_{i}$ at given $s$ using
$2s-3$ known terms of the moment expansion. Further we test the
correctness of so obtained $C_{n}^{s}$ from the comparison with the
remaining known terms in the moment expansion, which were not used to
determine $a_{i}$. The exact knowledge of $50$ moments enables one to
determine $C_{n}^{s}$ for $s\leq 11$. The coefficients $C_{m}^{s}$ we
have obtained in this way
are%
\begin{eqnarray*}
 C_{n}^{1} =\frac{1}{2n+1}, \\
 C_{n}^{2} =-\frac{n}{3}, \\
 C_{n}^{3} =\frac{1}{90}n(n+1)(14n+1), \\
 C_{n}^{4} =-\frac{1}{1890}n(n+1)(2n+3)(62n^{2}+3n-2), \\
 C_{n}^{5} =\frac{1}{37800}n(n+1)(n+2)(2n+3)(508n^{3}-84n^{2}-19n+15), \\
 C_{n}^{6} =-\frac{1}{3742200}%
n(n+1)(n+2)(2n+3)(2n+5)(10220n^{4}-6236n^{3} \\
 +1597n^{2}+737n-372), \\
 ...
\end{eqnarray*}%
We have also computed the coefficients $C_{n}^{7}$ and $C_{n}^{8}$, but
do not write them here to save space. The determination of higher
coefficients demands the knowledge of more moments of the end-to-end
distribution function and can be obtained in a similar way. After the
determination of the
coefficients $C_{n}^{s}$ we have checked the equivalence of (\ref%
{w12}) with (\ref{mom-exp}) to the corresponding order. Unfortunately, we
did not succeed in deriving the expression for the coefficients
$C_{n}^{s}$ for arbitrary $s$.

With known coefficients $C_{n}^{s}$ one can sum the series over $n$ in (\ref%
{w12}). Restricting ourselves to the leading order $s_{\max }=1$
($s_{\max }$ is the number of terms in the sum over $s$ in (\ref{w12}))
we obtain $G(k,N)$
\begin{equation*}
G_{\mathrm{r}}(k,N)=\frac{\sin \left( kN\right) }{kN},
\end{equation*}%
and the distribution function of the end-to-end distance
\begin{equation*}
G_{\mathrm{r}}(r,N)=\frac{1}{4\pi N^{2}}\delta (N-r).
\end{equation*}%
in the stiff rod limit. Taking into account the next term in the sum over
$s$ in Eq.(\ref{w12}) gives the correction to the stiff rod limit of the
end-to-end distribution function. Thus, the resummation of the moment
expansion of $G(k,p)$ according to (\ref{w12}) yields the short-chain
expansion for Kratky-Porod chain.

The above derivation does not allow to make claims on the convergence of
the short-chain expansion. The comparison of results of computations of quantities
under consideration (for example the scattering function or
deformation-force relation) for different $s_{max}$ gives a criterion
determining the quality of the short-chain expansion.

Notice that alternatively one can first carry out the inverse Laplace
transform of (\ref{mom-exp}), and then expand the moments in Taylor series
in powers of $N$. As a result one will arrive at double series similar to (%
\ref{w12}) with $1/p$ replaced by $N$. This makes more clear the meaning
of the resummation procedure. To leading order ($s_{\max }=1$) one
replaces the moments by its stiff rod behaviour, $\left\langle
R^{2n}\right\rangle =N^{2n} $. In the next to leading order ($s_{\max
}=2$) one takes into account the next-order corrections to the stiff rod
behaviour of all moments, and so on.

Due to the fact that the scattering function of the semiflexible polymer $%
S(k,N)$ is the inverse Laplace transform of $G(k,p)/p^{2}$ multiplied
with the factor $2/N$, the short-chain expansion of $G(k,p)$ enables one
to get the short-chain expansion of the scattering function. In fact,
Hermans and Ullman \cite{hermans/ullman52} derived the stiff rod limit of
the scattering function using the stiff rod limit of the moments.

In following subsections we will consider separately the short-chain
expansion of the end-to-end distribution function, the scattering
function, and the extension-force relation, which also can be obtained from $%
G(k,N)$.

\subsection{Distribution function of the end-to-end distance}

The series over $n$ in (\ref{w12}) for known coefficients $C_{n}^{s}$ can
be easily expressed through the derivatives of the geometric series.
Carrying out the inverse Laplace transform over $p$ gives the short-chain
expansion of $G(k,N)$. We have computed $G(k,N)$ by taking into account
eight terms in the sum over $s$ in (\ref{w12}) i.e. $s_{\max }=8$. To
save place
we present below the result for $s_{\max }=5$%
\begin{eqnarray}
\fl G_{5}(k,N)=G_{\mathrm{r}}(k,N)+\frac{1}{6}\,{\frac{\sin \left( kN\right) }{%
k}}-\frac{1}{6}N\,\cos \left( kN\right)  \nonumber \\
\fl +{\frac{1}{60}}\,{\frac{\sin \left( kN\right) N}{k}}-{\frac{7}{360}}%
\,k\sin \left( kN\right) {N}^{3}-{\frac{1}{60}}\,{N}^{2}\cos \left( kN\right)
\nonumber \\
\fl +{\frac{1}{630}}\,{\frac{\sin \left( kN\right) {N}^{2}}{k}}+{\frac{1}{5040}%
}\,k\sin \left( kN\right) {N}^{4}-{\frac{1}{630}}\,\cos \left( kN\right) {N}%
^{3}+{\frac{31}{15120}}\,\cos \left( kN\right) {N}^{5}{k}^{2}  \nonumber \\
\fl +{\frac{1}{5040}}\,{\frac{\sin \left( kN\right) {N}^{3}}{k}}-{\frac{1}{5600%
}}\,k\sin \left( kN\right) {N}^{5}+{\frac{127}{604800}}\,{k}^{3}\sin \left(
kN\right) {N}^{7}  \nonumber \\
\fl -{\frac{1}{5040}}\,\cos \left( kN\right)
{N}^{4}-{\frac{53}{151200}}\,\cos \left( kN\right) N^{6}{k}^{2}.
\label{gkN}
\end{eqnarray}%
The subscript at $G$ and at quantities below is $s_{\max }$. Since $%
G_{s}(k,N)$ tends to one for $k\rightarrow 0$, the end-to-end
distribution function is normalized. It is easily to see that the terms
in (\ref{gkN}) can be represented as derivatives of $\sin \left(
kN\right)/k$ with respect to $N$ i.e. of $NG_{\mathrm{r}}(k,N)$. The
short-chain expansion of the distribution function of the end-to-end
distance is then immediately obtained from (\ref{gkN}) as
\begin{eqnarray}
\fl \pi G_{5}(r,N)=\frac{1}{4r^{2}}\delta \left( N-r\right) +\frac{1}{12r}%
\delta \left( N-r\right) -\frac{1}{24}\delta ^{(1)}\left( N-r\right) +
\frac{1}{24}\delta \left( N-r\right)\nonumber \\
\fl -\frac{r}{30}\delta ^{(1)}\left(
N-r\right) -{\frac{7}{1440}}\,r^{2}\delta ^{(2)}\left( N-r\right) +
{\frac{2}{63}}\,r\delta \left( N-r\right) {\,}-{\frac{31}{1008}}%
r^{2}\,\delta ^{(1)}\left( N-r\right) \nonumber \\
\fl -{\frac{11}{1440}}\,r^{3}\delta
^{(2)}\left( N-r\right) -{\frac{31}{60480}}\,r^{4}\delta ^{(3)}\left(
N-r\right) + {\frac{5}{144}}\,r^{2}\delta \left( N-r\right) -
{\frac{37}{1008}}r\delta^{(1)}\left( N-r\right) \, \nonumber \\
\fl-{\frac{131}{11200}}r^{4}\delta ^{(2)}\left(
N-r\right) \,-{\frac{209}{151200}}r^{5}\delta ^{(3)}\left( N-r\right) \,-{%
\frac{127}{2419200}}\,r^{6}\delta ^{(4)}\left( N-r\right),
\label{grN}
\end{eqnarray}
where $\delta^{(k)}(x)$ denotes the kth derivative of the Dirac's delta-function.
Note that $G(r,N)$ is a distribution function with respect to $r$, while
the contour length $N$ is a parameter. This should be taken into account
in using Eq.(\ref{grN}) to compute different mean values, for example the
moments of the end-to-end distribution function of a short chain. Similar
to the corresponding expansion of the function $\delta_a(x)=1/(2\pi
a^2)^{1/2}\exp(-x^2/{2a^2})$ in powers of $a$, the short-chain expansion
of $G(r,N)$ cannot be interpreted as a Taylor series but is rather a
generalized Taylor expansion in space of distributions. The expansions
(\ref{gkN}-\ref{grN}) can be used to test approximative expressions of
the end-to-end distribution function. We did not succeed so far to
convert the short-chain expansion of the distribution function to a close
expression.

\subsection{The scattering function}

\begin{figure}[tbph]
\begin{center}
\includegraphics[clip,width=2.5in]{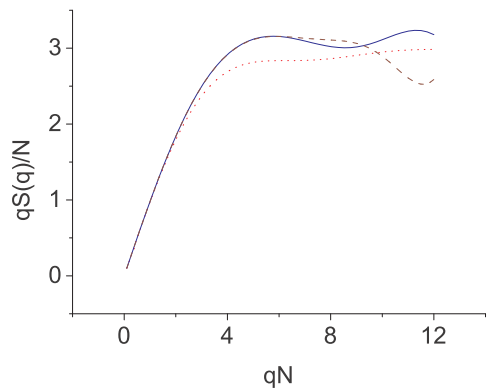}
\end{center}
\caption{The plot of $qS(q,N)$ for chain length $N=1.5$. Dots: stiff rod;
continuous: $s_{\max }=5$; dashes: $s_{\max }=8$. } \label{fig1}
\end{figure}

The scattering function of a semiflexible polymer is defined by
\begin{equation}
S(q,N)=\frac{2}{N}\int_{0}^{N}ds_{2}\int_{0}^{s_{2}}ds_{1}\left\langle \exp
(i\mathbf{q}(\mathbf{r}(s_{2})-\mathbf{r}(s_{1})))\right\rangle .
\label{w17}
\end{equation}%
Expressing $\mathbf{r}(s_{2})-\mathbf{r}(s_{1})$ in (\ref{w17}) through the
tangent vectors, $\mathbf{r}(s_{2})-\mathbf{r}(s_{1})=\int_{s_{1}}^{s_{2}}ds%
\mathbf{t}(s)$, and representing the average in (\ref{w17}) using the
formalism of the quantum rigid rotator yields that the scattering
function of the semiflexible polymer $S(q,N)$ is the inverse Laplace
transform of $G(q,p)/p^{2}$ multiplied with the factor $2/N$ \cite{SS02},%
\cite{stepanow04}. Thus, the short-chain expansion of $G(k,p)$ enables
one to derive in a straightforward way the short-chain expansion of the
scattering function. Taking into account the first four terms in the sum
over $s$ in (\ref{w12}) results in the following expression for the
scattering function
\begin{eqnarray*}
\fl S_{4}(x,N)/N=S_{\mathrm{r}}(x)/N+\frac{2}{3}{\frac{N}{{x}^{2}}}-{\frac{%
N\sin x}{{x}^{3}}}+\frac{1}{3}{\frac{N\cos x}{{x}^{2}}}+{\frac{7}{180}}\,{%
\frac{{N}^{2}\sin x}{x}}+\frac{6}{5}{\frac{{N}^{2}}{{x}^{4}}}- \nonumber \\
\fl {\frac{13}{15}}\,{\frac{{N}^{2}\sin x}{{x}^{3}}}+{\frac{4}{15}}\,{\frac{{N}%
^{2}\cos x}{{x}^{2}}}-\frac{6}{5}{\frac{{N}^{2}\cos x}{{x}^{4}}}-{\frac{32}{%
63}}\,{\frac{{N}^{3}}{{x}^{4}}}+{\frac{307}{7560}}\,{\frac{{N}^{3}\sin x}{x}}%
- \nonumber \\
\fl {\frac{125}{126}}\,{\frac{{N}^{3}\sin x}{{x}^{3}}}+3\,{\frac{{N}^{3}\sin x%
}{{x}^{5}}}+{\frac{31}{126}}\,{\frac{{N}^{3}\cos x}{{x}^{2}}}-{\frac{157}{63}%
}\,{\frac{{N}^{3}\cos x}{{x}^{4}}}-{\frac{31}{7560}}\,{N}^{3}\cos x,
\end{eqnarray*}%
where $x=qN$, and
\begin{equation*}
S_{\mathrm{r}}(x)/N=2\,{\frac{\cos \left( x\right) }{x^{2}}}-\frac{2}{x^{2}}%
\,+2\,{\frac{\mathrm{Si}\left( x\right) }{x}}
\end{equation*}%
is the scattering function of a stiff rod. The plot of the scattering
function multiplied by $q$ for different values of $s_{\max }$ is shown
in Fig.\ref{fig1}.
The accuracy of the computations is determined by the values of $x=qN$,
where curves corresponding to different $s_{\max }$ begin to diverge. The
Fig.\ref{fig1} shows that for $N=1.5$ the continuous and dashed
curves begin to diverge for $x\geq 9$.

\subsection{The extension-force relation}

\begin{figure}[tbph]
\begin{center}
\includegraphics[clip,width=2.5in]{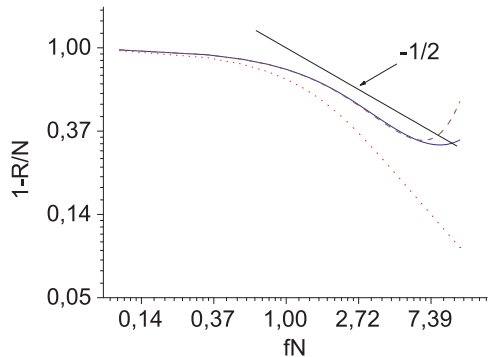}
\end{center}
\caption{Extension-force relation for chain length $N=1.1$. Dots: stiff
rod; continuous: $s_{\max }=4$; dashes: $s_{\max }=6$.}
\label{fig4}
\end{figure}

The partition function of a semiflexible polymer with one end fixed and
force $\mathbf{f}$ applied to the another end
\begin{equation}
Z(\mathbf{f},N)=\left\langle \exp (-\frac{\mathbf{f}}{k_{B}T}\int_{0}^{N}ds%
\mathbf{t}(s))\right\rangle   \label{w18}
\end{equation}%
can be expressed through the distribution function of the end-to-end
distance as follows
\begin{equation}
Z(\mathbf{f},N)=G(\mathbf{k}=-i\frac{\mathbf{f}}{k_{B}T},N).  \label{w19}
\end{equation}%
Using the definition of the free energy $F=-k_{B}T\ln Z(f,N)$ the
extension-force relation can be expressed through the partition function as
\begin{equation}
R=-\frac{\partial F}{\partial f}=k_{B}T\frac{\partial \ln Z(f,N)}{\partial f}%
.  \label{w20}
\end{equation}%
Thus, the short-chain expansion of the extension-force relation is
directly obtained from that of the Fourier transform of the distribution
function of the end-to-end distance (\ref{gkN}). Taking into account the
first $s_{\max }=4$ terms in the series in (\ref{w12}) results in the
following expression for the extension-force relation
\begin{eqnarray}
\fl \frac{R}{N} =-\frac{1}{x}+\coth x -\frac{1}{6}\,Nx-\frac{1}{6%
}N\coth x+\frac{1}{6}Nx\coth ^{2}x+{\frac{1}{90}}\,N^{2}\coth x\nonumber \\
\fl +\frac{1}{20}%
\,N^{2}x  -\frac{1}{36}N^{2}x^{2}\coth x
-{\frac{7}{180}}\,N^{2}x\coth ^{2}x+\frac{1}{36}N^{2}x^{2}\coth ^{3}x\nonumber \\
\fl -\frac{1}{1512}\,N^{3}\coth x  +{\frac{11}{1512}}\,N^{3}x^{2}\coth x-{\frac{19%
}{2520}}\,xN^{3}
 -{\frac{1}{120}}\,N^{3}x^{2}\coth ^{3}x\nonumber \\
\fl +{\frac{1}{840}}\,N^{3}x^{3}-{\frac{%
11}{1890}}\,N^{3}x^{3}\coth ^{2}x+{\frac{1}{216}}\,N^{3}x^{3}\coth ^{4}x+{%
\frac{11}{2520}}\,N^{3}x\coth ^{2}x,\,  \label{R-f}
\end{eqnarray}
where $x=fN$.
The first two terms on the right-hand side of (\ref{R-f})
gives the extension-force relation for stiff rod. The log-log plot of $1-R/N$
for different $s_{\max }$ and for stiff rod is shown in Fig.\ref{fig4}.
The slope for the stiff rod is $-1$, while the slope for the worm-like
chain approaches the value $-1/2$, which is the asymptotic result for
finite worm-like chain at large forces \cite{marko-siggia95}.
Fig.\ref{fig4} shows that the convergence of the short-chain expansion
for the extension-force relation is worse than that of the scattering
function.

\section{Conclusions}

To conclude, using the exact computation of a large number of moments of
the end-to-end distribution function $G(r,N)$ of the worm-like chain,
which are obtained from the representation of the distribution function
as a matrix element of the infinite order matrix
\cite{SS02},\cite{stepanow04}, we have established the analytical form of
the coefficients in Taylor expansions of the moments for short $N$. The
knowledge of these coefficients enabled us to resummate the moment
expansion of $G(r,N)$ by taking into account consecutively the deviations
of the moments from their stiff rod limit. Within this procedure we have
derived the short-chain expansion for the distribution function of the
end-to-end polymer distance, the scattering function, and the
extension-force relation, by taking into account the deviations of the
moments from their stiff rod limit to the seventh order in $N$. The
procedure can be extended to higher orders. The short-chain expansion
could be useful in studies of the behaviour of short polymers, where the
deviation from stiff rod is small.

\ack {The support from the Deutsche Forschungsgemeinschaft (SFB 418) is gratefully
acknowledged.}

\end{document}